\documentclass[aps,superscriptaddress,altaffilletter,tightenlines]{revtex4}

\begin{document}

\title{EVOLUTION OF A UNIVERSE FILLED WITH A CAUSAL VISCOUS FLUID}

\author{Luis P. Chimento and Alejandro S. Jakubi                \\
{\it Departamento de F\'{\i}sica, Facultad de Ciencias Exactas y Naturales, }\\
{\it Universidad de Buenos Aires, Ciudad  Universitaria,  Pabell\'{o}n  I, }\\
{\it 1428 Buenos Aires, Argentina.}}

\begin{abstract}
The behaviour of solutions to the Einstein equations with a causal viscous
fluid source is investigated. In this model we consider a spatially flat
Robertson-Walker metric, the bulk viscosity coefficient is related to the
energy density as $\zeta = \alpha \rho^{m}$, and the relaxation time is
given by $\zeta/\rho$. In the case $m = 1/2$ we find the exact solutions and
we verify whether they satisfy the energy conditions. Besides, we study
analytically the asymptotic stability of several families of solutions for
arbitrary $m$. We find that the qualitative asymptotic behaviour in the far
future is not altered by relaxation processes, but they change the behaviour
in the past, introducing singular instead of deflationary evolutions or
making the Universe bounce due to the violation of the energy conditions.
\end{abstract}
\maketitle
\section{Introduction}

It is believed that quantum effects played a fundamental role in the early
Universe. For instance, vacuum polarisation and particle production arise
from a quantum description of matter. It is known that both of them can be
modelled in terms of a classical bulk viscosity \cite{Hu}.

Cosmological models with a viscous fluid have been studied by several
authors. Some of the interesting subjects addressed by them were the effects
of viscous stresses on the avoidance of the initial singularity \cite{Mur},
the dissipation of a primordial anisotropy~\cite{Mis}, the production of
entropy \cite{Wei71}, and inflation and deflation \cite{Bar86}.

Recently Pavon et al. considered a homogeneous isotropic spatially-flat
universe filled with a causal viscous fluid whose bulk viscosity is related
to the energy density by the power law $\zeta \sim \rho^m$ \cite{Pav}.

In this paper we extend and improve the analysis of this cosmological model.
We present the set of equations that describe the model in section 2. In
section 3 we solve them when $m=1/2$ and give a detailed analysis of their
solutions.  In section 4 we study of stability of those asymptotic solutions
that appear for $m\ne 1/2$ in the noncausal model by means of the Lyapunov
method. The conclusions are stated in section 5.

\section{The Model}

In the case of the homogeneous, isotropic, spatially flat Robertson-Walker
metric 

\begin{equation}
ds^{2} = dt^{2} - a^{2}(t) ( dx^{2}_{1} + dx^{2}_{2} + dx^{2}_{3} ) 
\end{equation}

\noindent only the bulk viscosity needs to be considered. Thus we replace in
the Einstein equations the equilibrium pressure $p$ by an effective pressure 
\cite{Wei71} 
\begin{equation}
H^{2} = {\frac{1}{3}} \rho \qquad \dot H+ 3H^{2} = {\frac{1}{2}} (\rho - p -
\sigma ) 
\end{equation}

\noindent where $H$= $\dot a/a$, $^{\cdot }=d/dt$, $\rho $ is the energy
density, $\sigma $ is the viscous pressure, and we use units $c=8\pi G=1$.
As equation of state we take

\begin{equation}
\label{3}
p = ( \gamma -1 ) \rho  
\end{equation}

\noindent  with a constant adiabatic index $\gamma \ge 0$, and $\sigma $ has
the constitutive equation of a viscoelastic fluid

\begin{equation}
\sigma + \tau \dot\sigma= - 3 \zeta H 
\end{equation}

\noindent Here $\zeta \ge 0$ is the bulk viscosity coefficient and $\tau$ is
the bulk relaxation time and causality demands $\tau>0$. Following \cite
{Bel79} we choose

\begin{equation}
\zeta = \alpha \rho ^{m} \qquad \tau = \zeta /\rho 
\end{equation}

\noindent where $\alpha $ and $m$ are constants. Using Eqs. 2--5 we get

\begin{equation}
\label{7a}
{\frac \gamma 3}{\frac{|H|^r}{H_0^r}}\ddot H+\left( \epsilon+\gamma ^2{
\frac{|H|^r}{H_0^r}}\right) H\dot H+{\frac 32}\gamma \left(
\epsilon-{\frac{|H|^r}{H_0^r }}\right) H^3=0,\quad r\neq 0,\quad \gamma \neq 0
\end{equation}
\begin{equation}
\label{7b}
{\frac{\gamma _0}3}\ddot H+(\epsilon+\gamma \gamma _0)H \dot H +{\frac
32}(\epsilon\gamma -\gamma _0)H^3=0,\quad r=0
\end{equation}
\begin{equation}
\ddot H+{\frac{3^{1-r/2}}{\gamma _0}}|H|^{1-r}-{\frac 92}H^3=0,\quad r\neq
0,\quad \gamma =0
\end{equation}

\noindent
where $\epsilon\equiv {\rm sgn} H$, $r\equiv 2m-1$, $H_{0}\equiv (\gamma
/\gamma _{0})^{1/r}/ \sqrt3$ and $
\gamma _{0}\equiv \sqrt3\alpha$. We assume $\alpha >0$.

\section{Case $r= 0$}

The general solution of Eq. \ref{7b} for $H>0$ takes the following parametric form:

\begin{equation}
\label{8}
H(\eta )={\frac{{\sqrt {\gamma _0/ 3}} }{{1+\gamma \gamma _0}}}
\left( {Ae^{\lambda _+\eta }+Be^{\lambda _-\eta }} \right)^{1/ 2} \qquad
t(\eta )={\frac{ {\gamma _0/ 3}}{{1+\gamma \gamma _0} }} \int {\frac{{d\eta }
}{{H(\eta )}}}  
\end{equation}

\noindent where $\lambda _\pm$ are the roots of $\lambda ^2+\lambda +{\frac{{
\gamma_0 (\gamma -\gamma _0)} }{{(1+\gamma \gamma _0)^2}}}=0 $ and $A$, $B$
are arbitrary integration constants.

\subsection{ One-Parameter Solutions}

The one-parameter families of solutions arise when either A or B vanishes
and can be obtained explicitly:

\begin{equation}
\label{10a}
H_{\pm }(t)=\nu _{\pm }/\Delta t,\qquad \gamma \neq \gamma _0
\end{equation}
\begin{equation}
\label{10b}
H_{+}=D,\qquad H_{-}=\nu _0/\Delta t,\qquad \gamma =\gamma _0
\end{equation}
\begin{equation}
\nu _{\pm }={\frac{1+\gamma \gamma _0}{3(\gamma -\gamma _{0)}}}\left\{ 1\pm
\left[ 1-{\frac{4\gamma _0(\gamma -\gamma _0)}{(1+\gamma \gamma _0)^2}}
\right] \right\} ^{1/2}\qquad \nu _0={\frac 23}{\frac{\gamma _0}{1+\gamma
_0^2}}
\end{equation}

\noindent Thus, expanding solutions '--' are always Friedmann, but solutions
'+' are Friedmann for $\gamma >\gamma _{0}$, de Sitter for $\gamma =\gamma
_{0}$ and explosive for $\gamma <\gamma _{0}$.

\subsection{ Two-Parameter Families of Solutions}

When $AB \neq0$, $a(t)$ can be written in closed form in terms of known
functions only for some values of $\gamma $ and $\gamma _{0}$. However, in
general, we need to study the solution in the parametric form of Eq. \ref{8}
. We obtain the following classification for the two-parameter families of
solutions :

\noindent A. The evolution occurs between singularities, it reaches a
maximum and recollapses again. The leading behaviour near the singularities
is Friedmann, as \ref{10a}-- for $\gamma \neq \gamma _{0}$ or \ref{10b}-- if 
$\gamma =\gamma _{0}$ . These solutions have particle horizons ($0<\nu
_{-}<2/3$).

\noindent B. There is a bounce with an explosive singularity in the past and
its behaviour in the future is either:

1. asymptotically Friedmann, as \ref{10a}+ for $\gamma >\gamma _{0}$.

2. asymptotically de Sitter, as \ref{10b}+ for $\gamma =\gamma _{0}$.

3. divergent at finite times, with leading behaviour \ref{10a}+ for $\gamma
<\gamma _{0}$.

\noindent C. The evolution begins at a singularity with a Friedmann leading
behaviour, as \ref{10a}-- for $\gamma \neq \gamma _{0}$ or \ref{10b}-- if $
\gamma =\gamma _{0}$; and so they have also particle horizons. Then it
expands, and its behaviour in the future is like B.

\noindent D. The evolution begins at an explosive singularity and ends at a
big-crunch singularity.

\subsection{Energy Conditions}

\noindent $\bullet$ Dominant energy condition (DEC): $\rho \ge |p+\sigma|
\Leftrightarrow -3H^2\le\dot H \le 0$.

\noindent
DEC is violated part of the time in the following families: A about the
maximum (if $\nu _{-}\ge 1/3$); B1, about the bounce as well as for large
times (if $\nu _{+}>1/3$); near the singularity in C1, C2 (if $\nu _{0}<1/3$
) and C3 (if $\nu _{-}<1/3$); C3, near the "explosion". DEC is violated
always in families A (if $\nu _{-}<1/3$), B2 and B3.

\noindent $\bullet$ Strong energy condition (SEC): $\rho +3p+3\sigma \ge 0
\Leftrightarrow \dot H+H^{2}\le 0$.

\noindent
SEC is satisfied always in families A and C1 (if $\nu _{+}\le 1$). It is
violated part of the time in the families: B1, about the bounce (if $\nu
_{+}\le 1$); for large times in C1 (if $\nu _{+}>1$), C2 and C3. SEC is
violated always in families B1 (if $\nu _{+}>1$), B2 and B3.

\section{Case $r\neq 0$}

We investigate the asymptotical stability of behaviors that occur in the
noncausal model by means  of the Lyapunov method.

\subsection{ Stability of the de Sitter solution}

For the de Sitter solution $H=H_0$ we rewrite Eq. \ref{7a} as a "mechanical
system".

\begin{equation}
\label{18}
{\frac{d}{dt}} \left[{\frac{1 }{2}} \dot H^{2} + V(H)\right] = -3
\dot H^2 H\left[ \gamma + {\frac{1}{\gamma }} \left({\frac{H_0}{H}}
\right)^{r}\right]  
\end{equation}
\begin{equation}
\label{19a}
V(H) = - {\frac{9}{8}} H^{4} + {\frac{9}{2}} {\frac{H_{0}^{r}}{4-r
}} H^{4-r}\qquad r\neq 4  
\end{equation}
\begin{equation}
\label{19b}
V(H) = - {\frac{9}{8}} H^{4} + {\frac{9}{2}} H_{0}^{4} \ln {\frac{
H}{H}}_{0}\qquad r=4  
\end{equation}

\subsection{ Stability of the Asymptotically Friedmann solution}

For $r>0$ it is easy to check that Eq. \ref{7a} admits a solution whose
leading term is $2/(3\gamma t)$. To study its stability we make  the change
of variables $H = v(z)/t$, $t^{r} = z$. Then this equation takes the form

\begin{equation}
\label{21}
{\frac{d}{dz}} \left[{\frac{1}{2}} {v^{\prime}}^{2} + V(v,z) \right]
= - {\frac{3H^{r}_{0}}{r \gamma }} {v^{\prime}}^{2} v^{1-r} + O \left( {\frac{1
}{z^2}} \right)  
\end{equation}

\begin{equation}
\label{23a}V(v,z) = {\frac{9 H_{0}^r}{2 r^{2}}} \left[{\frac{v}{4-r}} - {
\frac{2}{3\gamma (3-r)}}\right] {\frac{v^{3-r}}{z}} + O\left( {\frac{1}{z^2}}
\right)  
\end{equation}
\begin{equation}
\label{23b}
V(v,z) = {\frac{1}{2}} H_{0}^{3} \left( v - {\frac{2}{3\gamma }}
\ln v \right) {\frac{1}{z}} + O\left( {\frac{1}{z^2}}\right) , \quad r= 4  
\end{equation}

\subsection{ Stability of exponential-like solutions}

In the case of the solution for $\gamma =0$ and $m\ne0$ \cite{Bar} 
\begin{equation}
a(t)=\exp\left[\frac{2m}{2m-1}w_M t^{\frac{2m-1}{2m}}\right] \qquad
w_M=\left(3^{m+1}m\alpha\right)^{-\frac{1}{2m}} 
\end{equation}

\noindent  we make the change of variables $H =\left(sw(s)/t\right)$, $
t^{r/(r+1)}=s$ and get

\begin{equation}
\label{26}
{\frac{d}{ds}} \left[{\frac{1}{2}} {w^{\prime}}^{2}+W(w,s)\right] =
- {\frac{3^{1-r/2}}{\gamma _{0}}} {\frac{r+1}{r}} {\frac{{w^{\prime}}^{2} }{
w^{r-1}}} s + O\left( {\frac{1}{s}} \right)  
\end{equation}

\noindent where now the potential to leading order in $1/s$ is

\begin{equation}
\label{27a}
W(w,s) = - {\frac{r+1}{r^{2}}}\left[{\frac{3^{1-r/2} w^{3-r}}{
(3-r)\gamma _{0}}} + {\frac{9(r+1)w^{4}}{8}} \right ], \quad r \ne 3  
\end{equation}

\begin{equation}
\label{27b}
W(w,s)=-4{\frac{3^{1-r/2}}{9\gamma _0}}\ln w-w^4,\qquad r=3
\end{equation}

\section{Conclusions}

When $r=0$, the splitting of the large time asymptotic behavior of solutions
in terms of sgn $(\gamma-\gamma_0$ ), closely resembles the classification
for the noncausal solutions. However, causality makes new families of
solutions appear as the bouncing ones and those which expand from a
singularity but recollapse in a finite time at another singularity. Most
singular solutions have particle horizons.

We demonstrate that there is no value of $r$ for which there is a stable
expanding de Sitter period in the far past. This supports strongly the
conclusion of that causality avoids the deflationary behavior proposed by
Barrow \cite{Bar}.

If $r<0$, a stable inflationary phase occurs in the far future for any $
\gamma>0$; the condition $\gamma=\gamma_0$ is required if $r=0$, and such a
behavior is unstable for $r>0$. We note that the noncausal model has no
stable de Sitter solution if $r=0$. If $\gamma=0$ and $m<0$, the same faster
than exponential expansion found by Barrow \cite{Bar} is asymptotically
stable.

If $r>0$, we find that relaxation effects do not alter the perfect fluid
behavior $a\sim t^{2/(3\gamma)}$ for $t\rightarrow\infty$. This arises
because the viscous pressure decays faster than the thermodynamical
pressure. However, if $r=0$ and $\gamma>\gamma_0$, both pressures decay
asymptotically as $t^{-2}$ and the exponent becomes $\nu_+$. The perfect
fluid behavior becomes unstable if $r<0$.

Large negative viscous pressures may arise in stable evolutions which avoid
an initial singularity or have a viscosity-driven inflationary stage. If $r=0
$, no two-parameter solution satisfies the energy conditions.

\end{document}